\documentclass[twocolumn,showpacs,pre,aps,amssymb,amsmath]{revtex4}
\usepackage{graphicx}
\usepackage{epsf}

\newcommand{\be}{\begin{equation}}
\newcommand{\ee}{\end{equation}}
\newcommand{\bd}{\begin{displaystyle}}
\newcommand{\ed}{\end{displaystyle}}
\begin{document}
\oddsidemargin=0cm
\topmargin=0cm
\title{Atomic fractals in cavity QED}
\author{S. V. Prants and V. Yu. Argonov}
\affiliation{Laboratory of Nonlinear Dynamical Systems,
V.I.Il'ichev Pacific Oceanological Institute of the Russian Academy of Sciences,
690041 Vladivostok, Russia}
\date{\today}
\begin{abstract}

We report a clear evidence of atomic fractals in the nonlinear motion of a
two-level atom in a standing-wave microcavity. Fractal-like
structures, typical for chaotic scattering, are numerically found
in the dependencies of outgoing positions and momenta of scattered atoms on their
ingoing values and in the
dependence of exit times of cold atoms on their initial momenta in the generic
semiclassical models of cavity QED (1) with atoms in a far-detuned
amplitude (phase)-modulated standing wave and (2) with coupled  
atomic external and internal degrees of freedom. 
Tiny interplay between all the degrees of freedom in the second model is 
responsible for trapping
atoms even in a very short microcavity. It may lead simultaneously, at least,
to two kinds of atomic fractals, a countable fractal (a set of initial momenta
generating separatrix-like atomic trajectories) and a seemingly uncountable
fractal with a set of momenta generating infinite walkings of atoms
inside the cavity.
\end{abstract}
\pacs{42.50.Vk, 05.45.Df}
\maketitle

1. Cavity quantum electrodynamics (QED) with cold atoms is a rapidly growing
field of atomic physics and quantum optics studying the atom-photon interaction
in cavities \cite{B}. Recent experiments \cite{HL,PF} succeeded in
exploration coupled external atomic center-of-mass, internal atomic, and field
dynamics under condition of strong-coupling between a single cold atom and a single-mode field
in a high-finesse optical microcavity. Methods to monitoring the atomic
motion in real time have been realized experimentally. They open a new way
to study the very foundations of the matter-field interaction and use
particles trapped within a cavity mode for quantum communications, to
monitoring of chemical and biological processes at the single-molecule scale
and for other purposes.

The basic model Hamiltonian of the interaction of a two-level atom with a
single-mode standing-wave field in an ideal cavity is given by
\begin{eqnarray}
\label{1}
\hat H=\frac{\hat p^2}{2m}+\frac{1}{2}\hbar\omega_a
\hat\sigma_z+\hbar\omega_f\left(\hat a^\dagger\hat a+\frac{1}{2}\right)
-\nonumber\\
-\hbar\Omega_0 f(t)(\hat a^\dagger\hat\sigma_-+\hat a\hat\sigma_+)\cos(k_f\hat x),
\end{eqnarray}
where $\hat x$ and $\hat p$ are the atomic position and momentum operators,
$\hat\sigma$ the atomic Pauli operators, $\hat a$ and $\hat a^\dagger$ the field-mode
operators. We incorporate a function of time $f(t)$ in (\ref{1}) to describe
a possible modulation of the standing-wave amplitude in the framework of the
basic Hamiltonian. The strongly coupled atom-field system (\ref{1}) is a highly
nonlinear one and is known to possess a rich variety of qualitatively different
dynamics including chaos. The full Hamiltonian (\ref{1}) for an atom,
to be placed in a far-detuned modulated standing-wave, can be reduced to an
effective non-autonomous Hamiltonian
of a ``sructureless'' atom with one and half degree of freedom which
is a paradigma for quantum chaos in atomic optics \cite{GSZ92,MR94,AG98, HH01}.
It has been shown in \cite{PS,PK} that semiclassical chaos in the sense of
extremal sensitivity to small changes in
initial conditions is possible without any modulation, i.e. with an autonomous
dynamical system with three degrees of freedom, the field, external, and
internal atomic ones. In this Letter we report with both
the models a clear evidence of a new property of the atom-field dynamics,
atomic fractals in cavity QED.
In the strong-coupling limit, one neglects dissipation in all the degrees of
freedom that may be justified by available microcavities with very large
quality factors $Q\gtrsim 10^6$ \cite{HL,PF} or/and operating with far-detuned
atoms. 

2. {\bf Scattering of atoms by a far-detuned modulated standing wave.}
Following a scheme of experiments on scattering of atoms by
light \cite{GRP},
suppose that a monokinetic atomic beam propagates at an angle to
the direction $y$ which is perpendicular to the standing-wave axis $x$.
The field is assumed to be uniform in the $y$-direction. In a reference frame
moving with a constant velocity in the $y$-direction there remains only
the transverse atomic motion  which can be considered
classically if the atomic momentum is much larger than the photon momentum.
So the atom is treated as a point particle subject to a position-dependent
force and is thus deflected. One measures atomic positions $x$ and momenta $p$ after passing
the interaction zone.
For sufficiently large detuning between the field and atomic frequencies,
one can adiabatically eliminate the excited-state amplitude of an atom
\cite{GSZ92} and derive a dimensionless effective Hamiltonian for
the center-of-mass motion
\be
H_{\mbox{\it eff}}=\frac{\omega_r p^2}{2}- \varepsilon(1-\sin\beta\tau)
\sin^2 x,
\label{5}
\end{equation}
where $x= k_f<\hat x>$ and $p=\ <\hat p>/\ \hbar k_f$ are the scaled expectation
values for the atomic position and momentum, $\omega_r=
\hbar k_f^2/m\Omega_0$ and $\beta$ the scaled recoil
and modulation frequencies,  $\varepsilon$ and  $\tau=\Omega_0 t$
the scaled well depth and time, respectively.

\begin{figure}
\includegraphics[width=0.45\textwidth,clip]{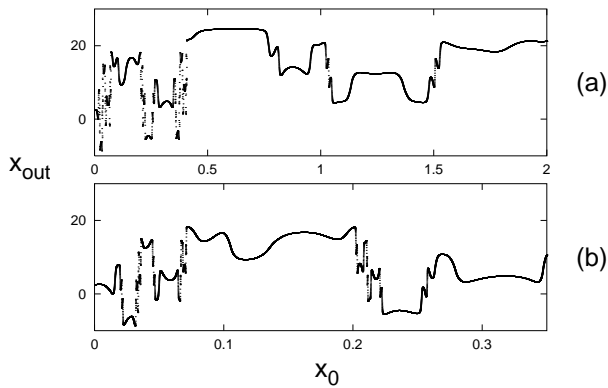}
\caption{The outgoing positions of far-detuned atoms $x_{\rm out}$ scattered by 
an amplitude-modulated standing wave as a function of initial positions $x_0$ in units
$k_f^{-1}$.}
\label{fig1}
\end{figure}

Solving the respective Hamilton equations we compute outgoing atomic
positions $x_{\rm out}$  at $\tau =500$ as a function of ingoing positions
$x_0$ with $\omega_r=0.001$, $\varepsilon =0.625$, and $\beta =0.07$.
Atoms in the beam are supposed to be initially in the ground state,
uniformly distributed along the $x$-axis
with the same value of the initial momentum $p_0=40.1$ and strongly detuned with
$\delta =(\omega_f-\omega_a)/
\Omega_0 =16$. Chaotic atomic
transversal motion in the  modulated standing wave results in a self-similar
structure of the scattering function which reveals itself under successive
magnifications (see FIG.~\ref{fig1}). Let us consider the initial interval
of $x_0$ (that may be identified with a cavity length) as a scattering region.
Chaos implies that there exist atomic trajectories that never escape this
region, and so there exist atoms never leaving the cavity. A set of initial
positions corresponding to nonescaping trajectories forms a fractal. We have
found singularities computing the time atoms need to go out off the
initial interval (to leave the cavity). Similar results with appearing 
fractals have been obtained with far-detuned atoms scattered by 
a phase-modulated standing wave. Fractal-like peculiarities should appear 
in standard experiments on scattering atoms at standing waves \cite{GRP}
in the case of their amplitude or phase modulations.

3. {\bf Fractals in the atom-field dynamics.} We wish to illustrate
a generic effect of fractals in cavity QED and use the full Hamiltonian
(\ref{1}) without any modulation, i.e. $f(t)=1$.
A comparatively large average number of photons is supposed in a single-mode
cavity to sustain atom-field Rabi oscillations. By using the following
expectation values: $x, \, p, \, z=\ <\hat\sigma_z>, \,
u=\ <\hat a^\dagger\hat\sigma_-+\hat a\hat\sigma_+>, $ and
$v= i<\hat a^\dagger\hat\sigma_--\hat a\hat\sigma_+>$, we translate
the Heisenberg equations with the Hamiltonian (\ref{1}) into the closed
nonlinear semiclassical system \cite{P02}
\be
\begin{array}{lll}
\dot x =\omega_r p, \, \dot p = -u\sin x, \, \dot u=\delta v, \,
\dot z= -2v\cos x,\\
\label{2}\\
\dot v=-\delta u + [(2N-1)z-3z^2/2+1/2]\cos x,
\end{array}
\end{equation}
where dot denotes a derivative with respect to $\tau$.
The dimensionless control parameters are the recoil frequency $\omega_r$,
the detuning $\delta$, and the number of excitations $N=<\hat a^\dagger\hat
a+(\hat\sigma_z+1)/2>$.
The integral of motion, $E=\omega_r p^2/2-u\cos x-\delta z/2$,
reflects the conservation of energy. At exact
resonance, $\delta=0$, the energy of the atom-photon interaction $u$ is
conserved, and the system (\ref{2})
becomes integrable with solutions describing regular atomic center-of-mass
motion in a potential well or over potential hills and atom-field Rabi
oscillations modulated by the translational motion \cite{PS}. Out of
resonance, the atomic translational motion is described by the equation for
nonlinear parametric pendulum, $\ddot x+\omega_r u(\tau)\sin x=0$, that has
been analytically shown \cite{P02} to produce weak chaos even in the case of
the simplest harmonic modulation $u(\tau)\sim\cos\Omega\tau$ caused by the
Rabi oscillations with a constant frequency $\Omega=\sqrt{\delta^2+4N}$. In
fact, the Rabi oscillations is an amplitude- and frequency-modulated signal
which may induce pronounced erratic motion of the atomic center-of-mass inside
the cavity. The motion is very complicated nearby the unperturped separatrix
of the nonlinear pendulum where the period of oscillations goes to infinity.
In this zone small changes in frequency, caused by respective small changes
in energy, may lead to dramatic changes in phase which are the ultimate reason
of exponential instability of atomic motion in a periodic standing wave.
The speculations above have been confirmed in our numerical experiments
\cite{PK,PS} where positive values of the maximal Lyapunov exponent have been
found in the following ranges of the control parameters: $\omega_r\gtrsim
10^{-3}$,
$N\lesssim 10^2$, and $\mid\delta\mid\ \lesssim 1.5$. These magnitudes seem to be reasonable
with available optical microcavities in which the strength of the atom-field
coupling may reach the values of the order of $\Omega_0\simeq 2\pi\cdot10^8$ Hz,
and the conditions of strong coupling are fulfilled for both the internal and
external degrees of freedom \cite{HL,PF}.
\begin{figure}
\includegraphics[width=0.30\textwidth,clip]{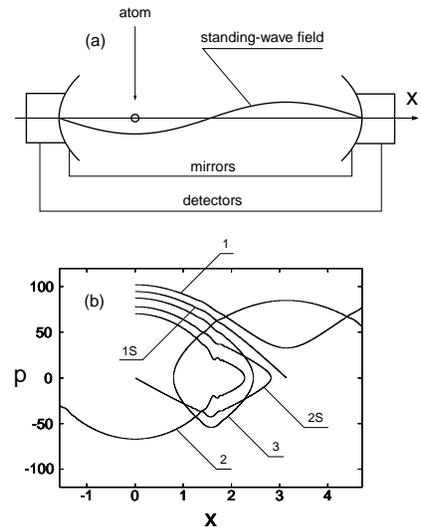}
\caption{(a) The schematic diagram showing a standing-wave microcavity with
detectors. (b) Sample atomic trajectories on the plane of atomic
momentum $p$ (in units $\hbar k_f$) and position $x$
(in units of $k_f^{-1}$).}
\label{fig2}
\end{figure}

Fractals in the dependencies $x_{\rm out}(x_0)$ and
$p_{\rm out}(p_0)$ with atoms being transversely injected into a 
stationary standing wave have been found with Eqs. (\ref{2}).
In FIG.~\ref{fig2}a we depict the scheme of another {\it gedanken} experiment
with a Fabry-Perot optical microcavity and two detectors that may lead to
another type of atomic fractals in cavity QED.
To avoid complications that are not essential to the
main theme of this work, we consider the cavity with only two standing-wave
lengths. Atoms, one by one, are placed at the point $x=0$ with different
initial values of the momentum $p_{0i}$ along the cavity axis. We measure the
time when an atom reaches one of the detectors, the exit time $T$, and
study the dependence $T(p_0)$ under the other equal initial conditions
imposed on the atom and the cavity field. At exact resonance,
optical potential coincides with the standing-wave structure, and
the analytical expression for the dependence in
question can be easily found to be the following:
$T(p_0)=\infty$ if $p_0<p_{\rm cr}/{\sqrt{2}}, \,
p=p_{\rm cr}$; $T(p_0)= {\rm F}[\arcsin(-1/K\sqrt{2}),K]/{\sqrt{\omega_r u_0}}$ 
if
$p_{\rm cr}/\sqrt{2}\leqslant p_0< p_{\rm cr}$; and
$T(p_0)=2{\rm F}[3\pi/4,K]/\omega_r p_0$ if $p_0>p_{\rm cr}$.
Here $K=p_0\sqrt{\omega_r/u_0}/2$ is the modulus of the first-order
incomplete elliptic integral {\rm F} and $p_{\rm cr}=2\sqrt{u_0/\omega_r}$
is the amplitude value of the atomic momentum on the separatrix. Atoms with
$p_0<p_{\rm cr}/\sqrt{2}$ are trapped in a potential well. Atoms
with $p_0>p_{\rm cr}$ fly through the cavity in one direction and are
registered by one of the detectors.

\begin{figure}
\includegraphics[width=0.45\textwidth,clip]{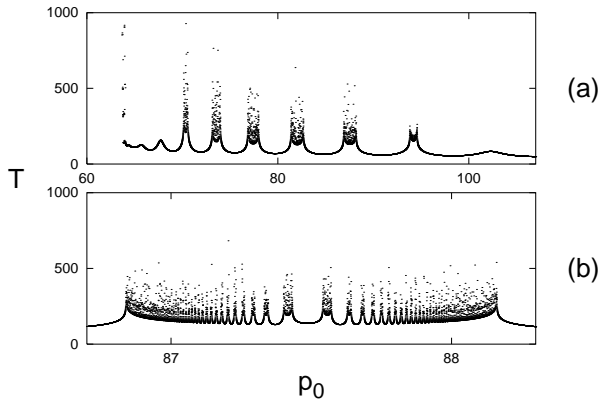}
\caption{Atomic fractal in the strongly coupled atom-field system with three 
degrees of freedom.
Exit time of atoms $T$  in units $\Omega_0^{-1}$ vs
the initial atomic momentum $p_0$ in units $\hbar k_f$.} 

\label{fig3}
\end{figure}

Out of resonance ($\delta\ne 0$), the atomic motion has been numerically found
\cite{PS,PK} and analytically proven  \cite{P02} to be chaotic with
positive values of the maximal Lyapunov exponent. FIG.~\ref{fig3} shows
the function $T(p_0)$ with the normalized detuning $\delta=0.4$, the
recoil frequency $\omega_r=10^{-3}$, the number of excitations $N=10$, and the
following initial conditions: $x_0=0$, $z_0=0$, and $u_0=v_0=2.17945$.
Atoms with comparatively small initial momentum, $p_0\lesssim 63$, cannot
reach the detectors because they are trapped in a well. With increasing
$p_0$, the exit time function demonstrates an intermittency of smooth
curves and complicated structures that cannot be resolved in principle, no
matter how large the magnification factor. 
FIG.~\ref{fig3}b shows magnification of the function for the small interval
$86.7\leqslant p_0\leqslant 88.3$. Further magnifications
reveals a self-similar fractal-like structure that is typical for
Hamiltonian systems with chaotic scattering \cite{Ott,HB}.
The length of the $T(p_0)$ function
$L= \sum_i \mid T_{i+1}-T_{i} \mid$ depends on the interval
$\epsilon$ of the partitioning of the momentum interval $p_0$.
We compute the length of the fractal curve in FIG.~\ref{fig3} to be
$L(\epsilon) \sim \epsilon^{-\gamma}$, where $\gamma \simeq 0.84 \pm 0.02$
is a fractal dimension simply connected with the Hausdorff dimension  $d =
1 +\gamma \simeq 1.84 \pm 0.02$.

The exit time
$T$, corresponding to both smooth and unresolved $\rho_0$ intervals, increases
with increasing the magnification factor. It follows that there exist atoms
never reaching the detectors in spite of the fact that they have no obvious
energy restrictions to leave the cavity. Tiny interplay between chaotic external
and internal dynamics prevents these atoms from leaving the cavity. The similar
phenomenon in chaotic scattering is known as {\it dynamical trapping}. Different
kinds of atomic trajectories, which are computed with the system (\ref{2}),
are shown in FIG.~\ref{fig2}b. A trajectory with the number $m$ transverses
the central node of the standing-wave before being detected $m$ times and is
called $m$-th trajectory. There are also special separatrix-like
$mS$-trajectories following which atoms in infinite time reach the
stationary points $x=\pm\pi n$ $(n=0, 1, 2, ...)$, $p=0$, transversing $m$ times the central node. These points are
the anti-nodes of the standing wave where the force acting on atoms is zero.
In difference from separatrix motion in the resonant system ($\delta=0$)
with the initial atomic momentum $p_{\rm cr}$, a detuned atom can
asymptotically reach one of the stationary points after transversing the
central node $m$ times. The smooth $p_0$ intervals in the first-order
structure in FIG.~\ref{fig3}a correspond to atoms transversing
once the central node and reaching the right detector. The unresolved singular
points in the first-order structure with $T=\infty$ at the border between the
smooth and unresolved $p_0$ intervals are generated by the
$1S$-trajectories. Analogously, the smooth and unresolved $p_0$ intervals
in the second-order structure in FIG.~\ref{fig3}b correspond to
the 2-nd order and the other trajectories, respectively, with singular points between them
corresponding to the $2S$-trajectories and so on.

There are two different mechanisms of generation of infinite detection times,
namely,
dynamical trapping with infinite oscillations ($m=\infty$) in a cavity and the
separatrix-like motion ($m\ne\infty$). The set of all initial momenta generating
the separatrix-like trajectories is a countable fractal. Each point in the set
can be specified as a vector in a Hilbert space with $m$ integer nonzero
components. One is able to prescribe to any unresolved interval of
$m$-th order structure a set with $m$ integers, where the first integer is a
number of a second-order structure to which trajectory under consideration
belongs in the first-order structure, the second integer is a number of a
third-order structure in the second-order structure mentioned above, and so on.
Such a number set is analogous to a directory tree address:
"$<$a subdirectory of the root directory$>$/$<$a subdirectory of the 2-nd
level$>$/$<$a subdirectory of the 3-rd level$>$/...". Unlike the separatrix fractal, the set
of all initial atomic momenta leading to dynamically trapped atoms with
$m=\infty$ seems to be uncountable.

\begin{figure}
\includegraphics[width=0.35\textwidth,clip]{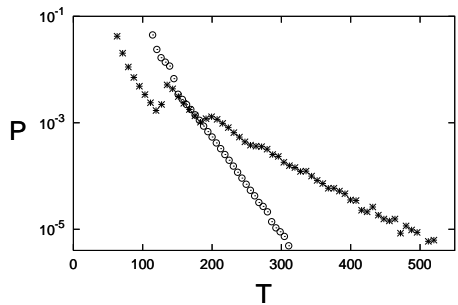}
\caption{Exit time distributions in the regimes of strong chaos with
$\delta=0.4$ (crosses) and weak chaos with $\delta=0.05$ (circles) with
the same initial conditions and the other control parameters, as 
in FIG.~\ref{fig3}.}
\label{fig4}
\end{figure}

The dependence of the maximal Lyapunov exponent of the set (\ref{2}) on the
detuning $\delta$ have been found almost the same as compared with a slightly
different version of (\ref{2}) considered in \cite{PS}. It shows a deep gorge
around $\delta=0$, maxima around $\mid\delta\mid\ \simeq 0.5$, and falls to zero
at $\mid\delta\mid\ \gtrsim 1.5$. We collect an exit time statistics, by counting atoms
with different initial momenta reaching the detectors, in the regimes of
comparatively strong chaos ($\delta=0.4$) and weak chaos ($\delta=0.05$).
The plots of the respective histograms of exit times are shown in
FIG.~\ref{fig4}. The probability distribution function $P(T)$ with almost
resonant atoms decays rapidly and demonstrates a single local maximum around
$T\simeq 140$. In the regime of strong chaos, $P(T)$ demonstrates a
few local peaks (the first one around $T \simeq 140$) and a long tail up to
$T\simeq 500$.

4. In summary, fractals in single-atom standing-wave cavity QED typically arise
in the nonlinear models of the atom-field interaction which are chaotic in the
semiclassical limit. It has been shown as for the non-autonomous Hamiltonian
system (\ref{5}) with one and half degree of freedom describing the nonlinear
motion of a ``structureless'' atom in a far detuned amplitude 
(phase)-modulated standing-wave field,  as for the autonomous system 
(\ref{2}) with three degrees of freedom
including internal and external atomic and field ones. The fractals
of the types to be found in this Letter should appear also in the center-of-mass motion
of an ion confined in a Paul trap and interacting with a laser field
\cite{W}. Different effective Hamiltonians have been introduced to describe
this interaction. The simplest one \cite{KB} is of a periodically driven
linear oscillator which is chaotic in the classical limit \cite{Z}.

We have considered the semiclassical chaotic scattering. In quantum dynamics 
there are no individual atomic trajectories. It implies lacking of 
singularities in scattering functions, but classical fractals should manifest 
themself in fluctuations of corresponding quantum S-matrices whose structures 
for generic quantum scattering models have been shown to be described by 
random matrices \cite{BS}. We plan to treat the quantum-classical correspondence 
in chaotic scattering of atoms by a standing wave in the future.

We thank an anonymous referee for valuable comments.
This work was supported by the Russian Foundation for Basic Research under Grant
Nos. 02-02-17796, 02-02-06840, and 02-02-06841.


\begin{thebibliography}{99}
\bibitem{B} {\it Cavity Quantum Electrodynamics, Advances in Atomic,
Molecular, and Optical Physics, Supplement 2}, ed. by P.R. Berman (Academic,
San-Diego, 1994); Phys. Scr. {\bf T76}, 1 (1998).
\bibitem{HL} C.J. Hood, T.W. Lynn, A.C. Doherty, A.S. Parkins, and H.J. Kimble,
Science {\bf 287}, 1447 (2000).
\bibitem{PF} P.W.H. Pinske, T. Fischer, P. Maunz, and G. Rempe, Nature (London)
{\bf 404}, 365 (2000).
\bibitem{GSZ92} R. Graham, M. Schlautmann, and P. Zoller, Phys. Rev. A {\bf 45},
19 (1992).
\bibitem{MR94} F.L. Moore, J.C. Robinson, C. Bharucha, P.E. Williams, and
M.G. Raizen,
Phys. Rev. Lett. {\bf 73}, 2974 (1994).
\bibitem{AG98} H. Ammann, R. Gray, I. Shvarchuck, and N. Christensen, Phys. Rev.
Lett. {\bf 80}, 4111 (1998).
\bibitem{HH01} W.K. Hensinger {\it et al},
Nature {\bf 412}, 52 (5 July, 2001).
\bibitem{PS} S.V. Prants and V.Yu. Sirotkin, Phys. Rev. A {\bf 64}, 033412 (2001).
\bibitem{PK} S.V. Prants and L.E. Kon'kov, JETP Letters {\bf 73}, 180 (2001)
[Pis'ma Zh. Eksp. Teor. Fiz. {\bf 73}, 200 (2001)].
\bibitem{GRP} P.L. Gould, G.A. Ruff, D.E. Pritchard, Phys. Rev. Lett. {\bf 56},
827 (1986); T. Sleator, T. Pfau, V. Balykin, O. Carnal, and J. Mlynek,
Phys. Rev. Lett. {\bf 68}, 1996 (1992).
\bibitem{P02} S.V. Prants, JETP Letters {\bf 75}, 63 (2002) [Pis'ma Zh. Eksp.
Teor. Fiz. {\bf 75}, 71 (2002)].
\bibitem{Ott} E. Ott, {\it Chaos in dynamical systems} (Cambridge
University Press, Cambridge, England, 1993).
\bibitem{HB} C.F. Hillermeier, R. Bl\"umel, and U. Smilansky, Phys. Rev. A
{\bf 45}, 3486 (1992).
\bibitem{W} D.J. Wineland {\it et al},
J. Res. Natl. Stand. Technol. {\bf 103}, 259 (1998).
\bibitem{KB} D.I. Kamenev and G.P. Berman, {\it Quantum chaos:
a harmonic oscillator in monochromatic wave} (Rinton Press, Inc., Prinston
New Jersey, USA, 2001).
\bibitem{Z} G.M. Zaslavsky, R.Z. Sagdeev, D.A. Usikov, and
A.A. Chernikov. {\it Weak Chaos and Quasiregular Patterns} (Cambridge
University Press, Cambridge, England, 1991).
\bibitem{BS} R. Bl\"umel, and U. Smilansky, Phys. Rev. Lett. {\bf 60}, 477 (1988).
\end{thebibliography}
\end{document}